\documentclass[10pt,twocolumn,twoside]{IEEEtran}
\IEEEoverridecommandlockouts                            

\usepackage{url}
\usepackage{cite, comment}
\usepackage{amsmath,amssymb,amsfonts}
\usepackage{algorithmic}
\usepackage{graphicx}
\usepackage{textcomp}
\usepackage{balance}
\usepackage{xcolor}
\usepackage[utf8]{inputenc}
\usepackage{optidef}

\def\BibTeX{{\rm B\kern-.05em{\sc i\kern-.025em b}\kern-.08em
    T\kern-.1667em\lower.7ex\hbox{E}\kern-.125emX}}
\usepackage{lipsum}

\title{\LARGE \bf 
Productive Curtailment in Agrivoltaic Systems under Flexible Interconnection Agreements
}

\author{Marcus Wu and Anna Stuhlmacher
\vspace{-0.5cm}
 \thanks{ 
M. Wu and A. Stuhlmacher are with the Department of Electrical and Computer Engineering, Michigan Technological University, Houghton, MI, USA. Emails: {hongyiwu, annastu@mtu.edu}}
}

\renewcommand{\baselinestretch}{.97}
\allowdisplaybreaks

\begin{document}
\begingroup
\allowdisplaybreaks

\maketitle

\begin{abstract} 
Flexible interconnection agreements are increasingly used to streamline the distributed generation interconnection process by limiting real power exports and avoiding costly grid upgrades. Agrivoltaic systems--solar photovoltaic (PV) panels installed over agricultural land--can provide added value under these agreements by adjusting the PV panels away from sun tracking while increasing the sunlight available to crops. This technical note investigates the operation of agrivoltaics under flexible interconnection limits and evaluates their impact on both PV energy production and crop outcomes. We formulate an optimization problem that determines the time-varying tilt of a single-axis tracking agrivoltaic system to maximize energy production subject to a real power export limit over an entire growing season. The resulting PV operating schedules are then used to evaluate PV energy production and crop yield. In a case study, we demonstrate that agrivoltaic systems can comply with flexible interconnection agreements through operational adjustments that improve crop yield, distinguishing them from conventional PV systems that rely solely on inverter curtailment.
\end{abstract}
\vspace{-0.5cm}
\begin{IEEEkeywords}
agrivoltaic systems, flexible interconnection agreements, food-energy nexus, optimization, solar photovoltaics
\end{IEEEkeywords}

\vspace{-0.5cm}
\section{Introduction}
The number of distributed energy resources connecting to the power grid is rapidly reshaping power system planning and operation. In 2025, solar photovoltaic (PV) installations accounted for more than half of all new generation capacity added in the U.S.~\cite{EIA}. This rapid growth of distributed generation has placed significant pressure on traditional interconnection processes, which rely on detailed utility studies to ensure that new generation can be safely accommodated (e.g., without violating voltage or thermal limits, or disrupting protection coordination), and to identify any required grid upgrades~\cite{DOEi2x}. 

To help address this challenge, many utilities now publish hosting capacity maps to streamline interconnection screening. Hosting capacity represents the amount of distributed generation that can be added to a given portion of the grid without requiring infrastructure upgrades (see~\cite{DOE_HC_database} for a database of hosting capacity maps). Methods for evaluating hosting capacity remain an active research area. For instance, \cite{Dubey2017} proposes a stochastic analysis to determine static hosting capacity limits and \cite{HC_fairness} formulates a dynamic hosting capacity problem which considers fairness in PV curtailment across space and time. 

In parallel, flexible interconnection agreements have emerged as a mechanism to enable more distributed generation connections within hosting capacity limits by agreeing to operational constraints such as real power curtailment during specific grid conditions or in coordination with energy storage. For example, the California Public Utilities Commission has enabled renewable energy systems to interconnect under Limited Generation Profiles, which specify a time-varying maximum export amount based on grid conditions~\cite{DOE_Flexible_DER_EV_2024}. While these agreements can reduce interconnection costs for the PV owner by avoiding paying for grid upgrades, there are tradeoffs due to lost energy production from curtailment. However, for agrivoltaic systems, these tradeoffs may be less pronounced, as reductions in grid injections can be achieved through PV panel adjustments that simultaneously support agricultural outcomes. 

Agrivoltaics involves placing elevated PV panels over agricultural land, allowing the same land to be used for both energy generation and crop production. Agrivoltaics can reduce competition between food and energy systems, as well as unlock additional benefits such as improved PV panel efficiency due to evaporative cooling from the crops below. Although still an emerging approach, agrivoltaic deployment is growing in the U.S., with over 50 sites used in conjunction with crop production, where around 70\% are research sites \cite{OpenEI_AgrivoltaicsMap}. A key characteristic of agrivoltaic systems is the potential tradeoff between crop outcomes (yield) and PV outcomes (power production). In systems with dynamic single- or dual-axis tracking PV system configurations, this tradeoff can be managed by adjusting panel orientation away from conventional sun tracking to increase photosynthetically active radiation (PAR, i.e., sunlight useful for plant growth) reaching the crops below, typically at the expense of PV power output. One commonly used, simplified example of this approach is anti-tracking, where the PV panels are deliberately positioned to be parallel with the sun’s rays during certain periods of the day to prioritize crop light exposure~\cite{STRT, GRUBBS2024114018}.

There is a growing body of research investigating agrivoltaic benefits and the operation of dynamic PV panels to realize these benefits. Prior studies have quantitatively analyzed these benefits, including improvements in land use efficiency, microclimate, water loss, and soil health (see~\cite{KoamiCoBenefits} for a comprehensive review). Several studies have examined the operation of dynamic PV panels to help achieve these benefits. Anti-tracking approaches have been explored in \cite{STRT, GRUBBS2024114018}, where \cite{STRT} analyzed the tradeoffs between PV energy and ground irradiance given different durations of anti-tracking operation with a single-axis PV system, and \cite{GRUBBS2024114018} identified critical time windows for anti-tracking. However, neither of these approaches use formal optimization methods. Very few papers have formulated optimization-based approaches for real-time dynamic PV panel operation. Ref.~\cite{Mignoni2025} minimizes deviations away from a prescribed shading trajectory for single-axis PV panels, but does not explicitly consider the impact of PV operation on the irradiance reaching the field. In~\cite{stuhlmacher2024optimizing}, the authors optimize the position of dual-axis PV panels to maximize energy generation over a day while meeting a minimum amount of PAR reaching the field. Similarly,  in~\cite{Bruno2025}, the authors maximize energy generation for single-axis PV over a growing season while meeting a predefined daily PAR threshold. To the best of our knowledge, no paper has examined agrivoltaic systems in the context of flexible interconnection agreements or formulated an optimal agrivoltaic operation problem subject to real power export limits. Under a flexible interconnection agreement, agrivoltaic systems can adjust PV panels away from sun tracking to comply with export limits instead of simply curtailing power, while simultaneously providing benefits to crops below and avoiding possible grid upgrade costs. 

The objective of this technical note is to demonstrate this potential benefit through a simple yet illustrative case study. We consider an agrivoltaic system with single-axis tracking panels operated over an entire growing season. We formulate an optimization problem that maximizes PV energy production given PV constraints, a flexible interconnection limit, and irradiance inputs. We solve for the PV panel tilt angle, which is used to determine the field shading and crop yield through a shading analysis and crop model, respectively. The contributions of this work are i) the formulation of the optimization-based framework for agrivoltaic operation under a flexible interconnection agreement, ii) the quantification of crop and PV outcomes for agrivoltaic systems under interconnection constraints, and iii) a discussion of implications for agrivoltaic deployment and interconnection practices. The rest of this note is organized as follows: Section~\ref{section: formulation} presents the formulation for a single-axis tracking agrivoltaic system subject to a flexible interconnection agreement. Section~\ref{section: case study} demonstrates the approach in a case study, and Section~\ref{section: discussion} concludes.

\section{Problem Formulation} \label{section: formulation}
This section describes the modeling, optimization, and evaluation framework used to assess agrivoltaic performance under a flexible interconnection agreement. Specifically, we formulate an optimization problem that solves for the time-varying tilt angle of a single-axis tracking PV system subject to the PV constraints and real power export limit. The resulting PV operating schedule is then used as an input to a shading analysis and crop model so that we can evaluate both the electricity production and crop yield over an entire growing season. We focus on single-axis tracking systems, as they represent the most common dynamic configuration in current utility-scale PV and agrivoltaic installations. The optimization problem is solved over the full growing season for all time periods $t=1..T$, where each period has a duration of $\Delta T$. Fig.~\ref{fig: block diagram} illustrates our approach linking the optimization, shading analysis, and crop model. 
\begin{figure}
    \centering
    \includegraphics[width=\linewidth]{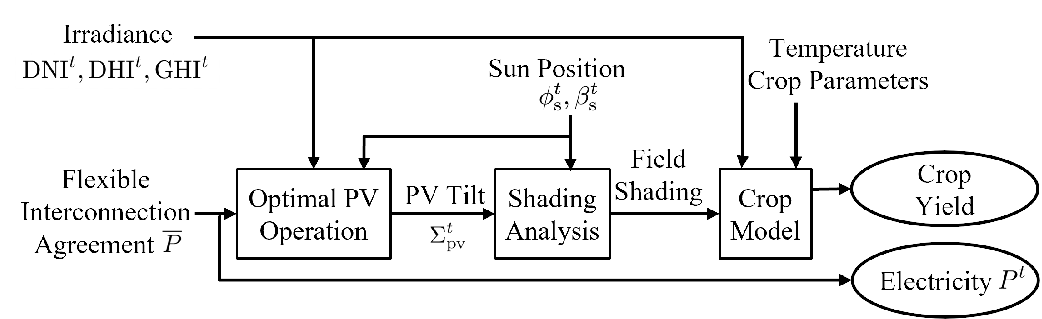}
    \vspace{-0.7cm}
    \caption{Block diagram of our approach to evaluate crop and PV power outcomes of an agrivoltaic system under a flexible interconnection agreement.}
    \vspace{-0.5cm}
    \label{fig: block diagram}
\end{figure}
In the remainder of this section, Section~\ref{subsection: PV model} presents the PV system model, Section~\ref{subsection: optimization} formulates the optimal operational problem, and Section~\ref{subsection: crop model} describes the shading and crop model approach.

\subsection{PV Model}\label{subsection: PV model}
We first present the model for the single-axis tracking PV system, where the position of the PV panels is described by the time-varying PV tilt angle $\Sigma_\text{pv}^t$ and the constant PV azimuth angle $\phi_\text{pv}$. We can control the PV tilt $\Sigma_\text{pv}^t$, which is constrained by its mechanical limits
\begin{align}
    \underline{\Sigma}_\text{pv} \leq \Sigma_\text{pv}^t \leq \overline{\Sigma}_\text{pv}, \label{eqn: tilt limits}
\end{align}
where $\underline{\Sigma}_\text{pv}$ and $\overline{\Sigma}_\text{pv}$ are the lower and upper limits, respectively.

The amount of power generated by the PV system depends on the amount of irradiance reaching the PV surface. The irradiance incident on the panels can be modeled as three components: direct $I_\text{direct}^t$, diffuse $I_\text{diffuse}^t$, and reflected $I_\text{reflect}^t$ for all time $t$~\cite{masters}
\begin{align}
    I_\text{direct}^t &= \text{DNI}^t\cdot\cos{\theta^t}, \label{eqn: direct irradiance}\\
    I_\text{diffuse}^t &= \text{DHI}^t\cdot \left(\frac{1+\cos{\Sigma_\text{pv}^t}}{2}\right),\label{eqn: diffuse irradiace}\\
    I_\text{reflect}^t &= \text{GHI}^t\cdot\rho\cdot\left(\frac{1-\cos{\Sigma_\text{pv}^t}}{2}\right), \label{eqn: reflected irradiance}
\end{align}
where $\rho$ is the ground reflectance and inputs $\text{DNI}^t$, $\text{DHI}^t$, and $\text{GHI}^t$ are the direct normal irradiance, diffuse horizontal irradiance, and global horizontal irradiance at time $t$. The angle of incidence $\theta^t$ is the relative angle between the direct beam and the normal vector of the PV panels at time $t$
\begin{align}
    \cos\theta^t =  \cos\beta_{\text{s}}^t \cos(\phi_\text{s}^t-\phi_\text{pv})\sin\Sigma_\text{pv}^t+\sin\beta_\text{s}^t\cos\Sigma_\text{pv}^t,\label{eqn: incidence angle}
\end{align}
where $\phi_\text{s}^t$ and $\beta_\text{s}^t$ are the sun azimuth and altitude. The power generated by the PV system at time $t$ is then 
\begin{align}
    P^t &= A_\text{pv}\cdot \eta_\text{pv}\cdot\left(I_{\text{direct}}^t+I_{\text{diffuse}}^t + I_\text{reflect}^t\right), \label{eqn: power}
\end{align}
where $A_\text{pv}$ is the total surface area of the PV panels in the array and $\eta_\text{pv}$ is the PV efficiency. 
The flexible interconnection agreement can either be fixed or dynamic over different time periods in accordance with grid conditions \cite{DOE_Flexible_DER_EV_2024}. In this work, without loss of generality, we define the flexible interconnection agreement as a static real power export limit $\overline{P}$
\begin{align}
    P^t &\leq \overline{P}. \label{eqn: export limit}
\end{align}
\subsection{Optimal PV Operation} 
\label{subsection: optimization}
The optimization problem solves for the time-varying PV tilt angle given inputs of solar irradiance, sun position, and the power export limit imposed by the interconnection agreement. The objective is to maximize PV energy production subject to the PV constraints and interconnection export limit. While crop-level targets could also be integrated into the optimization problem (e.g.,~\cite{stuhlmacher2024optimizing, Bruno2025}), we focus on PV energy maximization to isolate the impact of interconnection limits on agrivoltaic systems. The optimal operation problem is
\begin{maxi}|s|[2] 
{\mathbf{x}}
{ \sum_{t=1..T} \pi^{t} \, P^{t} \, \Delta T} {}{} 
\addConstraint{\eqref{eqn: tilt limits}-\eqref{eqn: export limit}\quad \forall\, t=1..T,}{}
\end{maxi}
where $\pi^t$ is the price of electricity at time $t$, and the decision variable vector $\mathbf{x}$ contains $\Sigma_\text{pv}^t$, $I_\text{direct}^t$, $I_\text{diffuse}^t$, $I_\text{reflect}^t$, $\theta^t$, and $P^t$ for all periods $t=1..T$. This formulation is nonconvex due to constraints \eqref{eqn: direct irradiance}-\eqref{eqn: incidence angle}, which makes the problem challenging to solve directly for seasonal time horizons.

To address this challenge, we reformulate and relax the problem into a second order cone program (SOCP) using an approach similar to~\cite{stuhlmacher2024optimizing} whose code is publicly available. The key reformulation differences are that we consider single-axis rather than dual-axis tracking, optimize over the growing season rather than a single day, include reflected irradiance, and include real power export limits. We first start by introducing auxiliary decision variables to rewrite the power, irradiance, and tilt angle decision variables relative to the results of a pre-specified sun-tracking algorithm that maximizes the collection of direct irradiance~\cite{masters}. For example, the PV tilt becomes
\begin{align}
    \Sigma_\text{pv}^t = \Sigma_\text{pv,st}^t + \delta\Sigma_\text{pv}^t,\label{eqn: sigma - ST and dev}
\end{align}
where $\Sigma_\text{pv,st}^t$ is the known sun-tracking PV tilt angle and $\delta\Sigma_\text{pv}^t$ is the deviation in PV tilt away from sun tracking. We define auxiliary variables $x^t := \cos{\delta\Sigma_\text{pv}^t,}$ and $y^t:=\sin{\delta\Sigma_\text{pv}^t}$, which introduces an additional constraint 
\begin{align}
    \left (x^t \right )^2 + \left(y^t \right)^2 = 1. \label{eqn: xy equality}
\end{align}
Using trigonometric identities to rewrite \eqref{eqn: direct irradiance}-\eqref{eqn: incidence angle} and relaxing \eqref{eqn: xy equality} into an inequality, we obtain an SOCP
\begin{maxi*}|s|[2] 
{\mathbf{x}}
{ \sum_{t=1..T} \pi^{t} \cdot (P_\text{st}^t + \delta P^{t}) \cdot \Delta T} {}{\tag{\textbf{Opt-Export}}} 
\addConstraint{\hspace*{-0.3cm}\delta I_{\text{direct}}^{t} = a_1^t x^{t} + a_2^t y^t - a_1^t }{}{\;\,\forall\, t=1..T}
\addConstraint{\hspace*{-0.3cm}\delta I_{\text{diffuse}}^{t} = b_1^t x^t + b_2^t y^t - b_1^t }{}{\;\,\forall\, t=1..T}
\addConstraint{\hspace*{-0.3cm}\delta I_{\text{reflect}}^{t} = c_1^t x^t + c_2^t y^t - c_1^t }{}{\;\,\forall\, t=1..T}
\addConstraint{\hspace*{-0.3cm}\delta P^{t} = A_\text{array}  \eta_\text{array} (\delta I_{\text{direct}}^{t} \!+\! \delta I_{\text{diffuse}}^{t}\! +\! \delta I_{\text{reflect}}^{t}) }{}{\;\, \forall\, t=1..T}
\addConstraint{\hspace*{-0.3cm}P_\text{st}^t + \delta P^t \leq \overline{P} }{}{\;\,\forall\, t=1..T}
\addConstraint{\hspace*{-0.3cm}d_3^t \leq  d_1^t x^t + d_2^t y^t \leq d_4^t }{}{\;\,\forall\, t=1..T}
\addConstraint{\hspace*{-0.3cm}\left (x^t \right )^2 + \left(y^t \right)^2 \leq 1, }{}{\;\,\forall\, t=1..T,}
\end{maxi*}
where parameters $a^t$, $b^t$, $c^t$, and $d^t$ are functions of the forecasted irradiance and sun-tracking position. As an example, plugging \eqref{eqn: sigma - ST and dev} into~\eqref{eqn: reflected irradiance} and applying cosine property $\cos{(a+b)} = \cos{a}\cos{b}-\sin{a}\sin{b}$, we get the coefficients for the deviation in reflected irradiance
\begin{align}
    \!c_1^t &= -\frac{1}{2}  \rho \cdot \text{GHI}^t \cdot  \cos{\Sigma_\text{st}^t}, \hspace{1.5em}
    c_2^t = \frac{1}{2}  \rho \cdot \text{GHI}^t \cdot  \sin{\Sigma_\text{st}^t}.
\end{align}
The decision variable $\mathbf{x}$ includes $x^t$, $y^t$, $\delta I_\text{direct}^t$, $\delta I_\text{diffuse}^t$, $\delta I_\text{reflect}^t$, and $\delta P^t$. The PV tilt can be recovered \textit{a posteriori}.

\subsection{Crop Model} \label{subsection: crop model}
Given the optimal PV tilt angle from (Opt-Export), we next compute the time-varying 
shading on the field using the approach described in~\cite{stuhlmacher2024optimizing}. This yields a field shading factor representing the percentage of the field that is shaded at time~$t$. The time-varying shading factor is then provided as an input into the crop model used to evaluate the yield. 

We use the EPIC crop model~\cite{williams1989epic}, a commonly used crop growth model that has unique crop-specific parameters. EPIC takes in temperature and irradiance reaching the field to calculate the yield given the heat unit accumulation, temperature stress, and leaf area index that vary across the growing season.  

\section{Case Study} \label{section: case study}
\subsection{Set Up} \label{subsection: set up}
To demonstrate the performance of an agrivoltaic system under a flexible interconnection agreement, we base our case study on publicly available data from an existing agrivoltaic research site from~\cite{OpenEI_AgrivoltaicsMap}: the Alliant Energy Solar Farm in Ames, Iowa~\cite{agrivoltaic_site}. 
We use Area 2's site information, which has a rated PV system capacity of approximately 150 kW under standard test conditions. The site is arranged in five rows, where each row contains three modules of 17 panels, with a 4-m edge-to-edge distance between rows. We estimate the panel dimension to be 2.576 m by 1.147 m and a 0.1-m spacing between modules in a row. We set $\rho$ to 0.2 and $\eta_\text{pv}$ to 20\%. We set the tilt limit to [-80$^\circ$, 80$^\circ$]. 
We vary the export limit, which is defined as a percentage of the PV system capacity. 

We select lettuce because it has a high relative yield under partial shading conditions~\cite{MARROU201354}, with the crop parameters pulled from~\cite{WinEPIC}. For solar data, we pulled the sun position from \texttt{pvlib}~\cite{anderson2023pvlib}, and the historical solar irradiance data from~\cite{NSRDB}. The solar data has a 10-minute resolution for a 60 day growing season from June 1st to July 31st, 2024. 

\subsection{Results} \label{subsection: results}
Table~\ref{tab:power_limit_results} evaluates energy production and crop yield under flexible interconnection limits ranging from 100\% (no curtailment) to 40\% of rated capacity. As the export limit decreases,  energy production decreases and crop yield increases. This occurs  because  the PV panels are shifted further away from sun tracking to curtail  power, which allows more accumulated PAR to reach the crops. In addition, the marginal reduction in energy increases at lower interconnection limits due to the increased duration and magnitude of curtailment. When the interconnection limit is reduced below 35\%, the optimization problem becomes infeasible without inverter curtailment, as panel reorientation alone cannot sufficiently reduce power output due to contributions from diffuse and reflected irradiance.  

It should be noted that the optimal solution will deviate away from the reference sun-tracking schedule $\Sigma_\text{pv}^t$ even when there are no active export limits. This occurs because the pre-specified sun-tracking algorithm in~\cite{masters} maximizes direct irradiance only, whereas our optimization formulation accounts for direct, diffuse, and reflected irradiance. As a result, the optimized schedule will achieve slightly higher energy generation. This result is related to the concept of backtracking in PV operation, where flatter panel positions in early morning and late evening are used to capture more diffuse irradiance when direct beam irradiance is low.

\begin{table}
\caption{Impact of Export Limits on PV Energy and Crop Yield}
\vspace{-0.4cm}
\begin{center}
\begin{tabular}{rrr}
\hline
\textbf{Power Limit (\%)} & \textbf{Energy (kWh)} & \textbf{Crop Yield (t/ha)} \\
\hline
100  & 73,501 & 1.874 \\
90  & 72,090 & 1.885 \\
80  & 68,534 & 1.937 \\
70  & 63,706 & 2.017 \\
60  & 57,583 & 2.129 \\
50  & 50,382 & 2.276 \\
40  & 42,257 & 2.455 \\
\hline
\end{tabular}
\label{tab:power_limit_results}
\end{center}
\vspace{-0.7cm}
\end{table}
 
To provide an illustrative example, we examine the power output of the PV system under varying weather conditions and interconnection limits in Fig.~\ref{fig: power trajectory}. Specifically,
we compare the actual power generation on a sunny and cloudy day using the optimal PV tilt angles from (Opt-Export). We observe that the PV power trajectory is reduced as the power export level shrinks. At higher limits (e.g., 80\%), the PV power output closely follows the unconstrained (100\%) profile and plateaus at the export limit in the middle of the day. However, we observe that PV energy production can sometimes exceed the interconnection limit, particularly during cloudy conditions at lower export limits (i.e., 40\% and 60\% interconnection limits on a cloudy day). This is because the relaxation of~\eqref{eqn: xy equality} is not always exact. In such cases, we set the tilt angle deviation by selecting the larger $\delta\Sigma_\text{pv}^t$ value from $x^t$ and $y^t$. During periods with more frequent power limit violations, we found that the direct irradiance is low relative to the diffuse and reflected components, which can make tilt deviations away from sun tracking capture more diffuse or reflected irradiance, resulting in a smaller feasible region. 

To quantify the frequency and magnitude of export limit violations, we calculate the average violation percentage over the entire growing season under different interconnection limits. We observe that the power limit violations are minimal (0-0.13\%) when the export limit is greater than 50\% and increase slightly (0.93-1.88\%) when the export limit is less than or equal to 50\%. In addition, violations are rare when wide tilt limits are allowed (e.g., $\pm$90$^\circ$) due to the increased operational flexibility of the PV system, although such ranges may not be realistic in all deployments. In practice, inverters could be used to supplement the mechanical curtailment to ensure strict compliance with flexible interconnection agreements. 

\begin{figure}
    \centering
    \includegraphics[width=\linewidth]{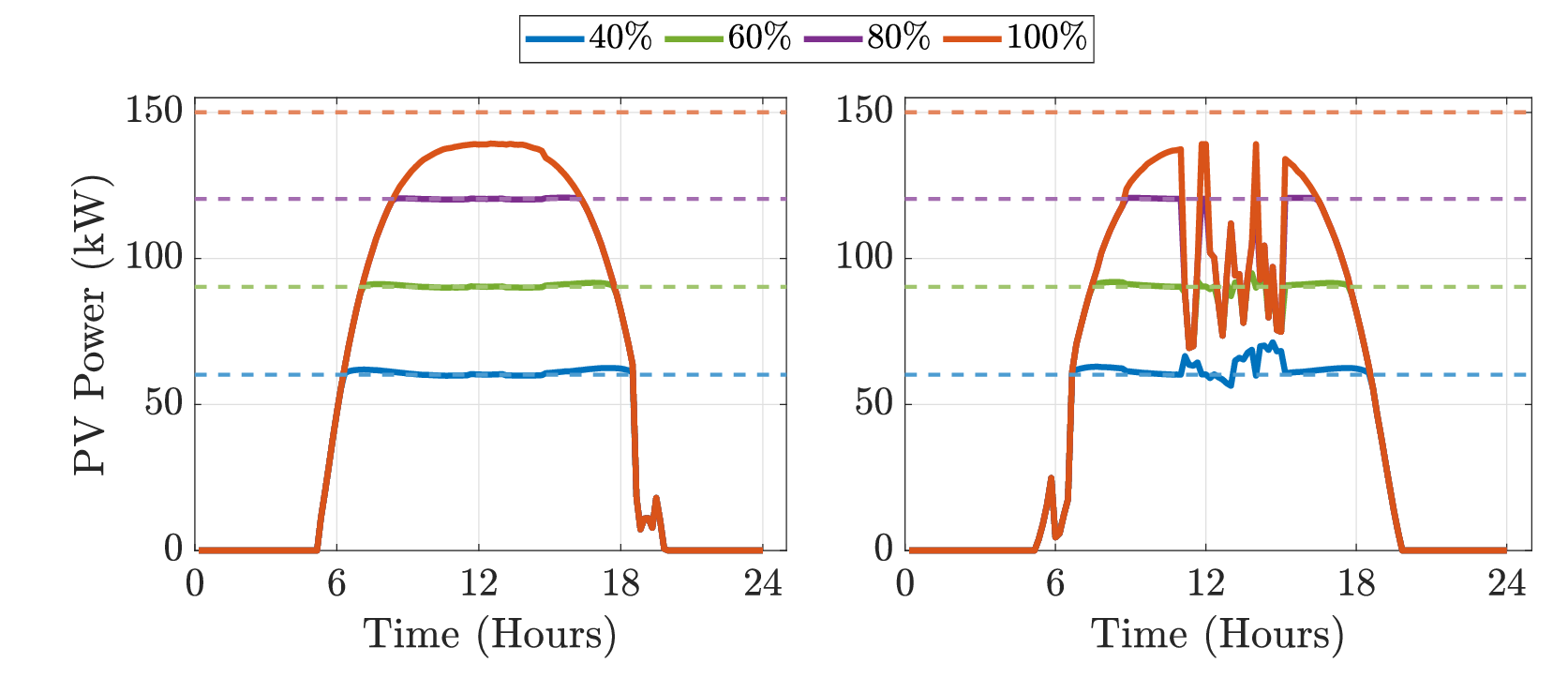}
    \vspace{-0.7cm}
    \caption{Example PV power profile on a clear day (July 14th, left) and a cloudy day (July 22nd, right) under different interconnection limits expressed as a percentage of rated capacity, with dashed lines of the same color indicating the corresponding real power limit $\overline{P}$.}
    \label{fig: power trajectory}
    \vspace{-0.5cm}
\end{figure}
\vspace{-0.35em}
\section{Discussion and Conclusions} \label{section: discussion}
In this technical note, we demonstrate how agrivoltaic systems can operate under flexible interconnection agreements by adjusting panel orientation to comply with real power export limits. This approach provides added value to the crops while  potentially avoiding grid upgrade costs associated with traditional interconnection processes.  Compared to conventional PV systems, allowing more PAR through to the crops below increases crop yield, making the tradeoff between reduced energy production and avoided grid upgrades less pronounced due to the agricultural value gained. In our case study, we find that when the interconnection power limit decreases from 100\% to 40\% of the nameplate capacity of the PV system, the crop yield increases by 31\%, highlighting a key distinction from inverter-based curtailment in PV-only systems.  

Much of the existing agrivoltaic research focuses primarily on agricultural and environmental outcomes. There has been significantly less attention focusing on the interactions between agrivoltaics and the power grid. However, as interest in agrivoltaics grows, it is becoming increasingly important to understand the potential impacts of agrivoltaic systems on the power grid, and vice versa. This is especially important given the differences in both design and operation of agrivoltaic systems compared to conventional PV-only installations. 

Future work will extend this framework to more detailed representations of both the PV system and underlying crops, including incorporating crop models directly into the optimal operation framework, imposing realistic limits on panel tilt rates, accounting for irradiance and temperature uncertainty, and exploring ways to improve the exactness of the convex relaxation. Together, these extensions will further clarify the role of agrivoltaic systems within an evolving power grid. 
\vspace{-0.25em}
\section{AI Usage Disclosure}
AI was used to assist in debugging the code. 
\vspace{-1em}
\bibliographystyle{ieeetr}
\bibliography{references}

\endgroup
\end{document}